# From Possibility to Precision in Macromolecular Ensemble Prediction

Stephanie A. Wankowicz[1*], Massimiliano Bonomi[2]

1) Molecular Physiology and Biophysics, Biochemistry, Center for Applied AI in Protein Dynamics, Center for Structural Biology, Vanderbilt University, Nashville, TN
2) Institut Pasteur, Université Paris Cité, CNRS UMR 3528, Computational Structural Biology Unit, Paris, France

*Correspondence: stephanie@wankowiczlab.com

## Abstract

Proteins and other macromolecules exist not in a single state but as dynamic ensembles of interconverting conformations, which are essential for catalysis, allosteric regulation, and molecular recognition. While AI-based structure predictors like AlphaFold have revolutionized static structure prediction, they are not yet capable of capturing conformational ensembles. Progress towards the next generation of AI models capable of ensemble prediction is currently limited by the lack of accurate, high-resolution ground truth ensembles at the scale required for training and validation. This is due to the fact that no single experimental technique can fully resolve the atomistic complexity of conformational landscapes, and fundamental challenges remain in defining, representing, comparing, and validating structural ensembles. Here, we outline the infrastructure and methodological advances needed to overcome these barriers. We highlight emerging strategies for integrating heterogeneous experimental data into unified ensemble encoding representations and how to leverage these new methodologies to build benchmarks and establish ensemble-specific validation protocols. Finally, we discuss how ensemble predictions will be an interactive cycle of experimental and computational innovation. Establishing this ecosystem will allow structural biology to move beyond static snapshots toward a dynamic understanding of molecular behavior that captures the full complexity of biological systems.

# Main

Biology is fundamentally a study of motion and change, where systems, from the environment to atoms, are never still. While biology relies on these motions to function, we tend to study biology through static representations, which restricts our understanding of many of its fundamental principles[1,2]. This limitation is evident in the study of macromolecules whose functions arise from conformational ensembles[3], yet are often interpreted as a single, static structure[4,5]. Macroscopic equilibrium properties like binding and stability emerge from a Boltzmann-weighted ensemble of conformations, where each microstate contributes to the partition function defining the system's thermodynamic behavior[3,6]. However, our structural interpretations of macromolecules and their macroscopic properties mainly stem from atomistic static models, often derived from cryo-electron microscopy (cryo-EM) and macromolecular crystallography, including X-ray, neutron, fiber, and electron diffraction (X-ray crystallography). While nuclear magnetic resonance (NMR) and molecular dynamics (MD) have provided critical insights into ensembles of conformational states, each method has its limitations. NMR is primarily constrained by system size and sensitivity[7], whereas MD is limited by force field accuracy and timescale accessible [8]. While static models have offered invaluable insights into structure and function, they represent a limited subset of the underlying conformational ensemble that drives biological activity. This overemphasis on static representations is perpetuated in protein structure prediction, limiting these algorithms' ability to predict protein functions[9–11]. Being able to computationally model and ultimately predict macromolecular conformational ensembles will have transformational impacts across biology by directly connecting structural models to macroscopic measurements, revealing disease mechanisms, evolutionary processes, and advancing next-generation therapeutics[12–14].

Recognizing the limitations in the static representations of macromolecules, the structural biology community has begun shifting its focus towards modeling conformational ensembles, details of which are often encoded in the very experimental data used to derive static structures[15–20]. This paradigm shift is also reshaping structure prediction, with growing numbers of efforts to predict conformational ensembles[21–24]. While incredibly promising, this shift brings new challenges: unlike static structures, where high-resolution models serve as widely accepted ground truths[25], conformational ensembles lack a single experimental technique capable of fully capturing the entire landscape at atomistic resolution, leaving the field without a definitive benchmark or gold standard for prediction validation. As a result, it remains difficult, if not impossible, for computational methods, including MD and structural ensemble prediction algorithms, to quantitatively validate predicted ensembles at atomic resolution across the full range of conformational states. Without a clear standard, the field faces foundational questions: What exactly are we trying to predict, and how will we know when we have succeeded?

# Protein Ensemble Prediction

The advancement in protein structure prediction achieved by AlphaFold2 highlights the transformative potential of curated biological data, rigorous benchmarks, evaluation metrics, and innovative AI approaches[11,26]. The Protein Data Bank (PDB), which primarily houses static macromolecular structural models, provided an extensive dataset for training predictive models. Simultaneously, the critical assessment of methods of protein structure prediction (CASP) community developed rigorous benchmarks and evaluation metrics for assessing structural predictions[27–29]. Without both of these contributions, it would have been impossible for the AlphaFold breakthrough of static protein structural prediction to occur. To replicate breakthroughs in conformational ensembles prediction, it is imperative to establish equivalent datasets, benchmarks, and evaluation tools, enabling innovations in encoding, architecture, and training. We see four key issues we need to address to achieve AlphaFold-like conformational ensemble predictions:

1) Structural ensembles are defined inconsistently across disciplines.

2) No single experimental technique alone can fully capture structural ensembles with the accuracy and precision needed to reflect their behavior in vivo.

3) Experimental approaches for determining ensembles face major challenges, including ensemble averaging, data sparsity, and intrinsic measurement errors.

4) Standardized representations, comparison metrics, and uncertainty quantification for structural ensembles are still lacking.

Here, we identify critical problems limiting our ability to capture macromolecular dynamics and propose actionable strategies to overcome them, calling on the experimental and computational structural biology communities to collaboratively advance experimental methods and predictive modeling beyond static snapshots toward ensemble-based representations.

# Defining Macromolecular Structural Ensembles

In statistical thermodynamics, conformational ensembles provide a framework to connect microscopic structural fluctuations to macroscopic properties. Even within its native, folded state, macromolecules exist as an ensemble of hierarchy-related substates, ranging from bond vibrations to localized side-chain rotamer flips to loop fluctuations to large domain rearrangements (**Figure 1**)[30,31]. The conformational ensemble encapsulates a protein's full range of states under a given set of conditions, with each state described by its Boltzmann factor reflecting its thermodynamic probability at equilibrium[3,32–35]. Collectively, these microstates define the protein's partition function ($Z=\sum_i e^{-\beta E_i}$; Z: partition function, $e^{-\beta E_i}$: Boltzmann probability density of i-th microstate; $E_i$: energy of i-th microstate), which governs macroscopic observable properties. Yet, the practical challenge lies in inferring these Boltzmann weights from incomplete, heterogeneous data, as we discuss below. Further, to relate these ensembles to function, we must also understand how the distribution of states' probabilities changes upon perturbation, such as ligand binding, mutation, or changes in environmental factors such as pH, temperature, or crowding[36,37].

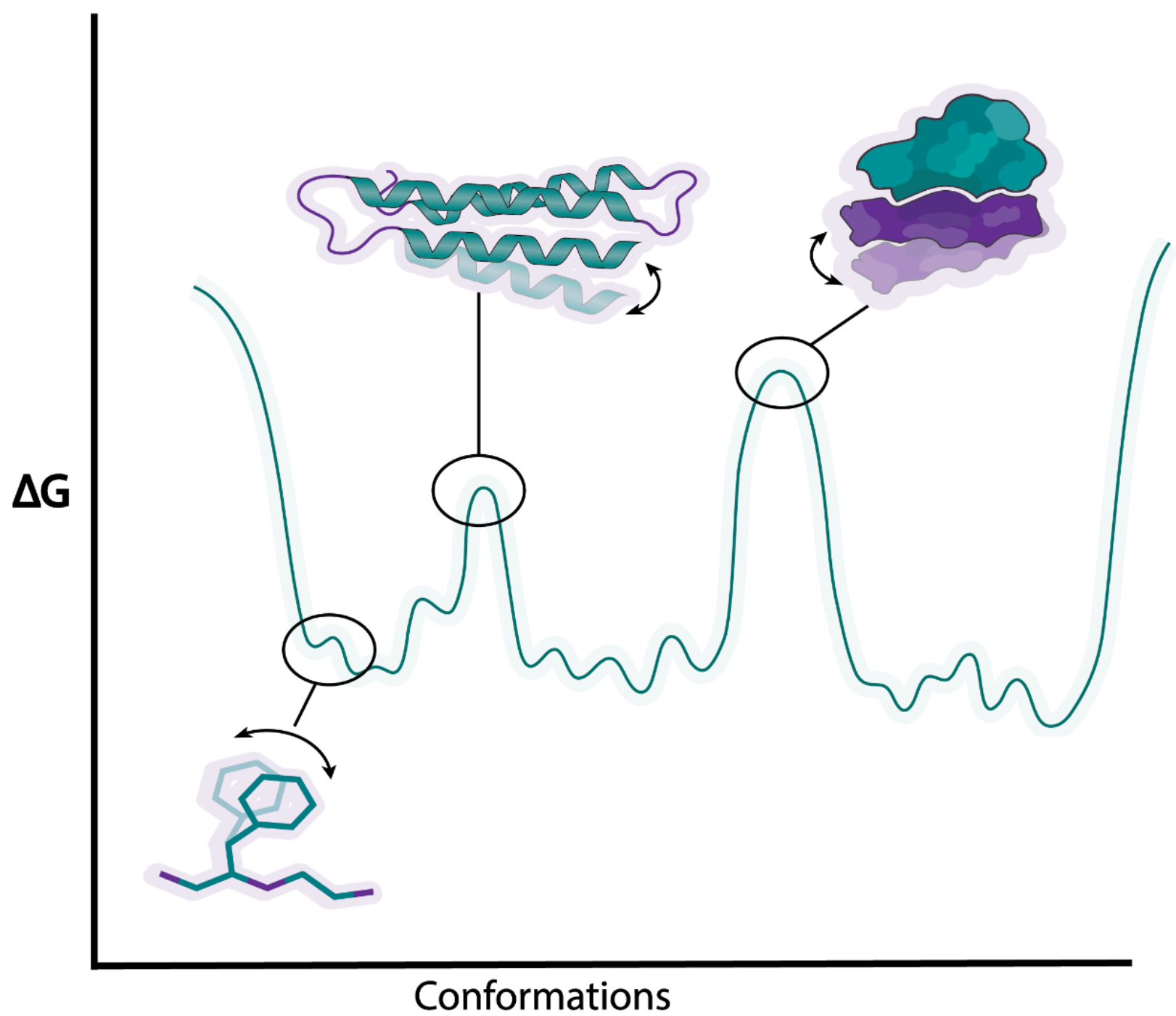

**Figure 1.** The hierarchical heterogeneity that makes up conformational ensembles, spanning from atomic vibrations, side-chain fluctuations, loop rearrangements, and large-scale domain motions, is naturally illustrated by the protein folding free energy landscape, where local valleys and peaks correspond to the vast ensembles of protein conformations. All changes within the conformational ensemble alter the shape of the free energy landscape and can have profound functional consequences.

Although some experimental methods and prediction algorithms often aim to represent conformational ensembles, they typically capture only a limited number of discrete macrostates, groups of microstates sharing similar structural or energetic properties, such as an active kinase. This oversimplification masks the full complexity of the ensemble, hindering our ability to accurately understand how the continuum of microstates contributes to observed macroscopic behavior[38]. Since every microstate influences the partition function, it is vital to consider the entire ensemble, despite the challenges in modeling or measurement.

One striking example of the complexity of determining conformational ensembles comes from the ribosome[39,40]. The ribosome's macro-structural states, such as in the rotated or elongation factor-bound states, are thermodynamically stabilized, with each conformation governed by a balance of enthalpic and entropic contributions[41–44]. Even minor fluctuations in the ribosome's protein and RNA components can shift the equilibrium among these states, influencing overall translation speed, fidelity, and responsiveness to external factors[39,40]. Oversimplifying the ribosome's thermodynamic ensemble by considering only macrostates overlooks key details that can influence evolutionary dynamics, lead to mistranslation, and affect drug design strategies.
The ribosome represents but one complex example of the ensemble nature of macromolecules. Other examples include transporters cycle between metastable states, with thermodynamic fluctuations driving function, as seen in GPCRs[45–47], multiple conformational changes enzymes have during their catalytic cycles, often populating rare but essential transient states that regulate chemical steps[48,49]. Further, ligand binding can reshape ensembles, as in nuclear receptors, where conformational changes toggle the receptor between transcriptionally active and repressed states[50]. Finally, intrinsically disordered regions (IDRs) highlight ensembles without a dominant folded structure, often enabling context-dependent interactions with diverse binding partners[51–53]. The diversity of systems and biological questions underscores the importance of modeling and predicting conformational ensembles.

## How do we generate ground truth ensembles?

The examples above underscore the potential of integrating experimental methods to unravel how conformational ensembles govern biological function. The link between ribosomal ensembles and their properties and functions was derived from many techniques and datasets, emphasizing that no single experimental or computational method can fully capture a structural ensemble, compelling us to move beyond the one technique, one structure, one solution mindset[54]. Constructing a conformational ensemble requires two key elements: identifying the conformation of each state at an atomistic level of uncertainty, and quantifying its contribution to the partition function.

First, we need to expand the algorithms pioneered by the integrative structural biology community to unite statistical analyses on multiple pieces of data from the same data types[55–57] (**Figure 2A**). Different experimental structural biology techniques provide unique information critical to capture a macromolecular conformational ensemble, but each has different limitations. Cryo-EM and X-ray crystallography can reveal high-resolution atomic structures, but are restricted due to their frozen state and/or crystalline environment. In contrast, NMR provides structural information averaged over multiple conformational states that interconvert more rapidly than the timescale of the NMR measurement. As a result, separating this averaged data into distinct conformational states and determining their relative populations remains a major challenge[58]. Spectroscopy approaches like FRET or double electron electron resonance (DEER) help detect rare states, though the sparsity of the data prevents the ability to resolve detailed macrostates fully[59]. Combining these concepts can include cryo-EM and X-ray, which could give us atomistic information on multiplicity of states, with local dynamics supported by NMR, and large macromolecular motions can be provided by spectroscopy techniques, SAXS, or Atomic Force Microscopy (AFM)[60–62].

Equally important is the use of a rigorous statistical structural biology framework to extract maximal information on conformational ensembles from any one technique. For example, by mining the wealth of structural experimental data in the PDB, analyzing multi-temperature or fragment-screening X-ray datasets, or integrating heterogeneous particle populations in cryo-EM, to reveal rare or otherwise hidden states[20,63–66](**Figure 2B**). All of this information can be greatly supported by incorporating computational approaches, such as MD or structure predictions, as discussed below.

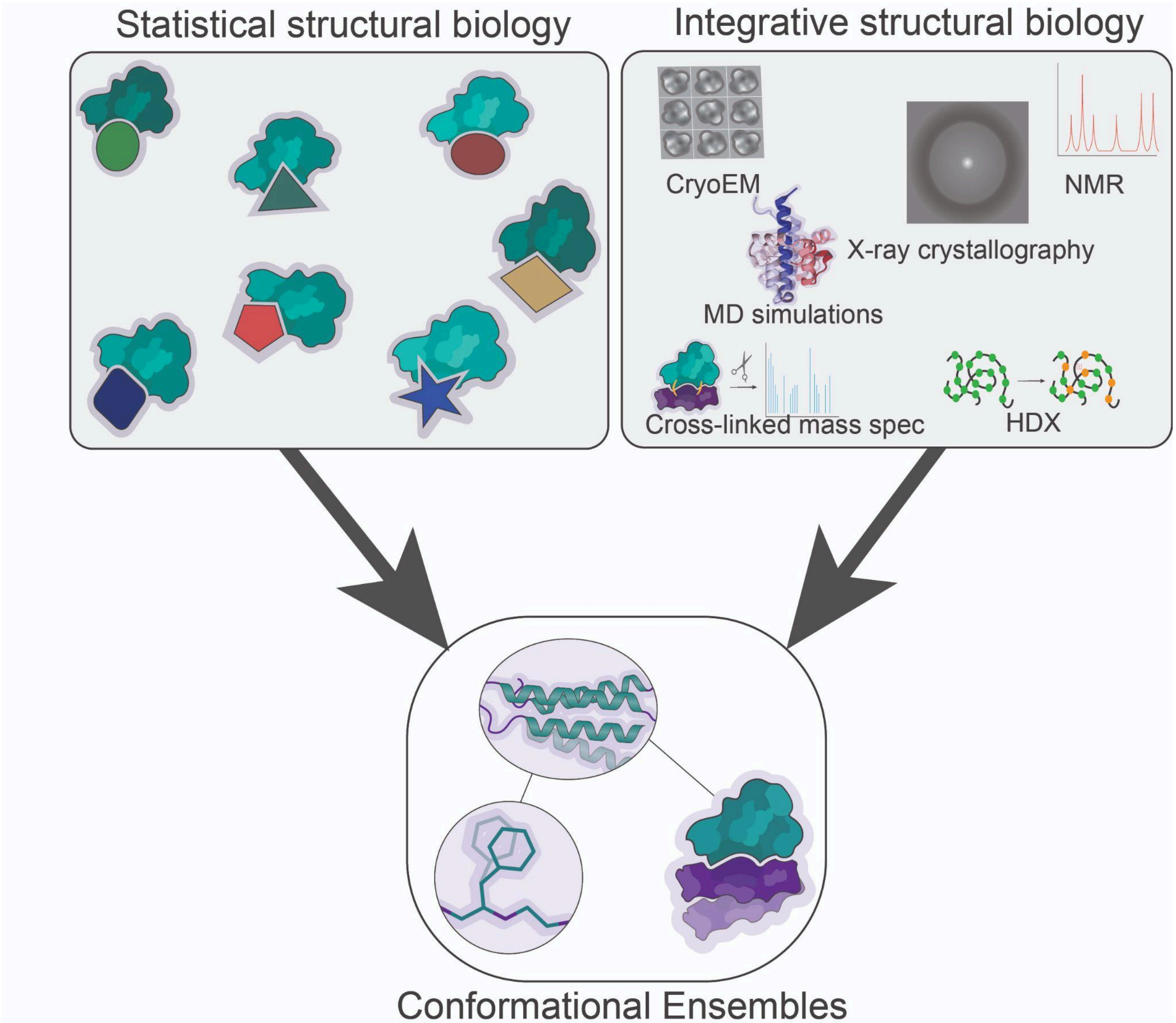

**Figure 2.** Strategies for generating gold-standard conformational ensemble datasets.
(A) Statistical methods applied to multiple datasets from the same technique (e.g., pseudoensembles in crystallography, heterogeneous cryo-EM particle populations, or replicate NMR measurements) can disentangle hidden states and quantify their relative populations. (B) Integrative approaches that combine distinct experimental modalities such as high-resolution cryo-EM or crystallography for atomistic states, NMR for local dynamics, and FRET, DEER, SAXS, or AFM for large-scale motions—provide complementary information across scales of heterogeneity. Together, these statistical and integrative strategies enable robust ensemble modeling that captures both common and rare conformational states.

We also must improve our ability to harness currently underutilized raw experimental data. Most experimental observables reflect time and ensemble averages over millions of conformations; thus, only conformational ensemble models accurately represent the experimental measurements[5]. However, most structural modeling algorithms only tend to model the most likely conformation, discarding the rich heterogeneity in the raw experimental data that reflects a broader conformational ensemble[67,68]. While new algorithms have begun to push the field beyond modeling a singular state, they still likely only capture a fraction of the accessible states[16,69,70]. Exciting new progress in determining conformational ensembles directly from cryo-EM particle stacks and using the often discarded diffuse scattering from X-ray crystallography can bring even more information about conformational ensembles in structural biology data into light[71–7374]. Continuous algorithmic development to model these "hidden" signals will provide a more comprehensive picture of macromolecular ensembles.

Reliable approaches are also needed to determine the statistical weights of conformations in a manner consistent with available data from both experiments and theory. For example, in cryo-EM, a key challenge is accurately fitting atomic models into continuously varying density maps while assigning appropriate Boltzmann probabilities that reflect the underlying energetic landscape[15]. While MD is often used to bridge these gaps[74], current biomolecular force fields have notable limitations, and the sampling problem presents an immense computational burden, particularly for large conformational changes, such as a domain movement[75–78]. Enhanced sampling techniques such as metadynamics or replica exchange could help fill missing conformational states, primarily if experimental data can anchor metastable states[18,79]. Yet scaling these approaches to obtain a complete structural ensemble across many biological systems remains a formidable challenge, requiring advances in high-throughput computation and potential neural network-based force fields[80].

Further, compositional heterogeneity, derived from ligand occupancy, stoichiometry, or assembly state, must be considered alongside conformational heterogeneity when weighting states[81]. Because most structural techniques are ensemble averaged, compositional and conformational variability are conflated; however, there are initial strategies to work on these. In cryo-EM, a normalized local filter can quantify per-voxel variance relative to local signal, highlighting regions where compositional differences dominate[82,83]. In crystallography, PanDDA event maps reveal low-occupancy ligands that might otherwise be mistaken for conformational disorder[84]; however, there are still significant challenges in disentangling the conformational and compositional heterogeneity outside of the binding site. Finally, time-resolved techniques or integration with orthogonal data sources such as native mass spectrometry provide an additional layer of validation, enabling ensemble models that explicitly separate composition from conformation[85,86].

Finally, we must continue advancing hardware and software technologies to enhance our capacity for collecting and analyzing high-quality structural data at scale. The challenge is twofold: efficiently generating the vast datasets needed to populate conformational ensembles using techniques we know can obtain these conformational ensembles, and designing the computational pipelines to integrate and analyze them in real-time. Examples include real-time cryo-EM data analysis and ongoing efforts to automate and streamline high-throughput X-ray data processing[87–90]. Innovations such as robotics for crystal harvesting, machine learning-assisted data evaluation[91,92], and real-time quality metrics offer promising steps forward. Integrating these noisy datasets with non-structural data, such as protein-protein interaction networks or genomic data, could further improve ensemble determination by uncovering hidden states[93,94].

# Experimental Challenges of Averaging, Sparsity, Errors, and Encoding

While collecting vast amounts of structural data at high-throughput scales holds great promise for uncovering conformational ensembles, accurately defining these ensembles remains challenging due to the ill-posed nature of the underlying inverse problem. Structural biology data typically capture ensemble-averaged measurements, resulting in an underdetermined scenario where multiple conformational ensemble models can equally fit the observed data (**Figure 3**)[95].

Furthermore, experimental noise inherently increases uncertainty and reduces the precision and reliability of reconstructed ensembles, with each experimental technique introducing distinct noise profiles. Within single techniques, new approaches explicitly account for technique-specific noise, for example, Maximum Entropy reweighting techniques that integrate cryo-EM with MD ensembles, can better separate genuine conformational heterogeneity from experimental uncertainty[18]. Other methods, such as variational autoencoders, have demonstrated potential for capturing continuous conformational states from cryo-EM data. Yet, their effectiveness can be limited by noise generated from different datasets, hindering generalizability across datasets[15,96].

Building on these approaches, a growing set of Bayesian inference frameworks addresses sparsity, averaging, and measurement error more broadly, opening the door to principled integration of data across multiple experimental modalities[78,97–101]. Such integrative approaches leverage the complementary strengths of different modalities, such as NMR or diffuse scattering, to identify correlated motion that can be connected to atomistic information in X-ray or cryo-EM data. Ultimately, the full potential of integrative structural biology hinges on developing sophisticated computational methods that accurately reweight conformational states based on diverse layers of experimental evidence.

Integrating techniques becomes more complex as proteins frequently exhibit differential dynamics across domains, with rigid cores coexisting alongside highly flexible loops or intrinsically disordered extensions. This heterogeneity poses additional challenges because of the assumptions underpinning many ensemble modeling approaches and comparison metrics. For example, global metrics such as root mean square distance (RMSD) or Jensen–Shannon divergence can overweight highly dynamic regions while underreporting subtle but functionally important motions in structured domains. Similarly, Bayesian reweighting or maximum entropy approaches must account for spatially varying uncertainty, as one domain may be well constrained by experimental data while another domain remains underdetermined.

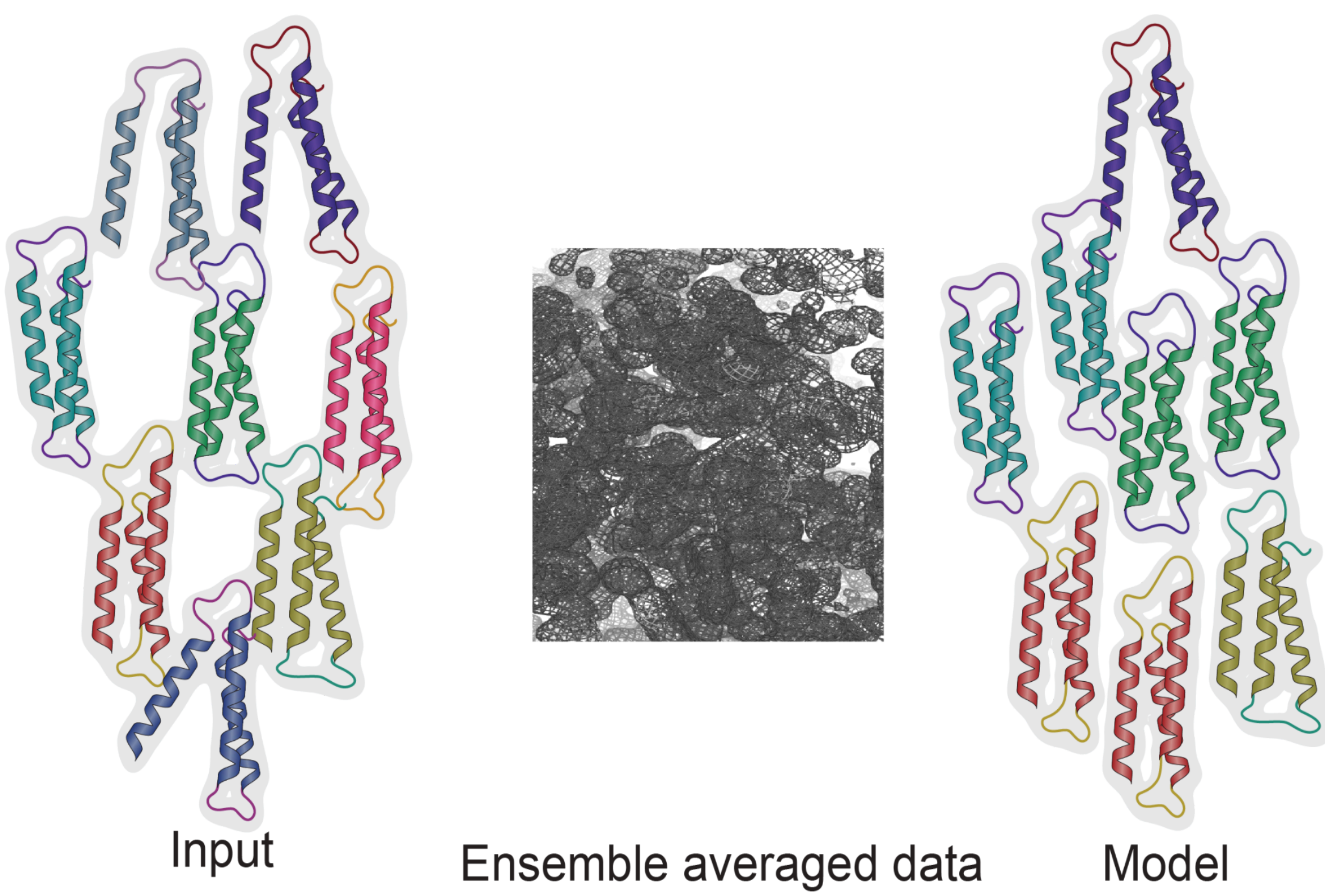

**Figure 3.** Modeling a conformational ensemble from a single structural modality is an ill-posed problem. For example, say you have eight different ‘true’ input conformations that comprise the experimental data (input). All of this data gets collapsed into ensemble-averaged data. While the model captures much of the structural ensemble, it overpopulates some conformations (i.e. green) and misses others (i.e. grey). However, it collectively explains much of the underlying experimental data.

Another complication is that inter-domain coupling can blur the boundary between local and global heterogeneity. Motions in one domain may allosterically influence the conformational distribution of another, leading to complex cross-correlations that are difficult to capture with independent models. Cryo-EM 3D variability analysis often highlights this uneven distribution of motion, but translating those observations into consistent ensemble encodings requires hierarchical representations that can treat domains differently while maintaining a unified partition function[20].

These challenges underscore that the encoding framework of ensembles often introduces limitations. The majority of structural data uses static PDBx/mmCIF models, which, while highly effective in encoding fixed atomic coordinates, fail to represent conformational ensembles adequately[102]. The current mechanisms to encode different conformational states within the format, including alternative locations (altlocs), B-factors, and multimodel encoding, are insufficient for fully capturing the variety of states that make up an ensemble[68,103]. Altlocs typically represent a parsimonious model, B-factors or atomic displacement factors mix genuine structural fluctuations with experimental noise or lattice movements[68], and multimodel encoding lacks crucial weighting information needed to represent conformational probabilities accurately[70].

The encoding of conformational ensembles falls along the spectrum from the maximum parsimony principle, which aims to explain the underlying data using the fewest parameters[16,104], or the maximum entropy principle, which seeks to create the least biased model distribution consistent with the underlying data by maximizing the Shannon entropy. Within experimental data modeling, maximum parsimony models, such as multiconformer models, allow for more direct encoding of occupancy or weights of different conformations, but are less expressive with subtle anharmonic motions and are limited in backbone movement. Maximum entropy methods can be advantageous in fitting very low occupancy states and picking up more subtle anharmonic motion[78,105], but the model complexity necessitates additional methods to extract biologically meaningful insights, such as identifying metastable states and assigning population weights. Other ensemble methods aim to explain the data collectively, but require re-weighting to be translated into maximum entropy models[70,106]. Regardless of the method used, the experimental data can be overfit if uncertainties are neglected or poorly modeled. Determining the best method for modeling and encoding ensembles, and how to accurately compare different methodologies, remains a fundamental challenge.

There is a possibility of taking the best from both approaches, though more expressive encoding of conformational ensembles in the PDBx/mmCIF format, including hierarchical information to represent relationships among structural states and separating conformational from compositional heterogeneity to improve interpretability[102]. Future work must include integration and encoding across experimental datasets, such as multiple cryo-EM maps. However, while these encoding approaches facilitate mapping data into a unified embedding space, they may also obscure signals due to uncertainties or low signal-to-noise ratios inherent in some techniques. Robust methodologies to compare and evaluate the effectiveness of different conformational ensemble encodings in resolving the inverse problem and translating results into biologically meaningful insights are critical, as they will also enable the rapid creation and testing of new encoding strategies.

# Comparing and Evaluating Structural Ensembles

In addition to encoding, we must establish rigorous frameworks for comparing conformational ensembles. While we have methods to compare coarse-grained ensemble-averaged descriptors of discrete-state heterogeneity, such as in SAXS or FRET[107], we still lack this ability at an atomistic level. For single static structures, evaluating the agreement between structures often involves root-mean-square deviation (RMSD), which measures the average atomic distance between two structures after alignment. Alternatively, specific structural features (e.g., dihedral angles) can be compared to experimental data, which is often represented by a sharply peaked posterior probability around the expected value[108]. However, accurately representing conformational ensembles requires shifting toward probability distribution functions (PDFs), enabling more physically realistic comparisons to experimental data. MD methods leverage symmetric divergence measures, such as Jensen-Shannon or Jeffrey's divergence, for quantitative PDF comparisons[109]. However, because PDFs are calculated from low-dimensional projections, a single PDF can map onto several structurally distinct, and potentially non-physical, ensembles. Furthermore, metrics based solely on PDFs are thus likely to encounter challenges when integrated into loss or optimization functions, reminiscent of initial difficulties faced with RMSD-based evaluations in CASP, leading to the development of alternative metrics, such as the Local Distance Difference Test (LDDT)[29]. More work is required to devise projection strategies that conserve the defining features of high-dimensional conformational spaces while potentially embedding relevant physical properties in their low-dimensional representations or other embedding strategies similar to those used with protein language models[110].

Alternatively, structural ensembles can be assessed based on their fit to experimental data. Several methodologies have been developed to compute forward models from MD simulations, which allow configurations to be constrained by experimental observables. Back-calculating intensities, maps, or raw particle stacks provides an additional means of evaluating ensemble accuracy[111,112]. However, multiple conformational states may satisfy the same experimental data, raising questions about how to determine the most physically meaningful representation. Considerable work is also needed to develop the best methods to compare conformational ensembles to single or multiple pieces of experimental data.

A central question in ensemble determination is evaluating whether the states and estimated weights of conformers are "good enough". The answer ultimately depends on the intended use of the ensemble. If the goal is primarily to validate and compare with experimental data, it is crucial to ensure accurate determination of the weights of conformers that contribute significantly to the measured observables. This can be particularly challenging when observables depend nonlinearly on structural properties. For example, in measurements like NMR nuclear Overhauser effect (NOEs) or FRET where signal scales with inverse sixth power of the distance ($1/r^6$), even sparsely populated conformers with short interatomic distances can significantly contribute to ensemble averaged observables, and their weights must therefore be determined with high precision[113].

In contrast, for functional applications, such as assessing the effect of a mutation or ligand binding, the critical issue is whether the inferred populations resolve differences above the level of statistical uncertainty, which can be quantified using bootstrap or jackknife resampling. Finally, when ensembles are linked to specific biological functions, the challenge becomes first identifying which structural states are responsible for specific functions and ensuring that their populations are accurately estimated.

# Integrating ML and MD to model conformational ensembles

While experimental data can collectively provide ground truth data to determine conformational ensembles, modeling these states cannot be done without computational methods, including MD and ML-based methods. While MD inherently attempts to model conformational ensembles, it is limited by the accuracy of the force fields and the timescales accessible in unbiased MD simulations[8]. A new generation of MD force fields, including machine-learned models, are beginning to enable atomistic simulations at quantum chemistry accuracy and coarse-grained potentials that retain atomistic predictive power[114–116], allowing longer and more accurate simulations. On the other hand, enhanced-sampling strategies can help overcome sampling limitations, but often require methodological expertise for each system, limiting automation and transferability[117]. Recent machine learning approaches have addressed these issues and improved enhanced sampling strategies, helping to accelerate MD exploration of conformational space[118].

However, the future is one in which generative ML-driven techniques alone will enable generating structural ensembles with accuracy comparable to atomistic MD with more biologically relevant time scales. Boltzmann generators first demonstrate this by using normalizing flows to generate unbiased conformations from the equilibrium state distribution[119]. More recently, generative diffusion models have taken over the static structure prediction space[120,121]. Using similar approaches, but often training with static structure, MD simulations, and in some cases, thermodynamic information, models have begun to predict pieces of conformational ensembles[21,24,122–124]. Notably, additional models were shown to efficiently determine the structure of intrinsically disordered proteins and regions[125,126].

Despite their promise, current ensemble prediction methods remain far from reliable. Work manipulating AlphaFold multiple sequence alignments (MSAs) has shown that structural diversity can be induced, but these approaches lack any notion of the relative probabilities of the resulting conformations[22,127]. They often generate thermodynamically unstable or improbable states, and it remains unclear when these methods fail or how to detect over-conditioning and template leakage, particularly in the absence of ground-truth ensembles for comparison[128]. Still, these efforts provide a foundation for building multimodal, generative AI-driven frameworks that incorporate experimental data at scale. Recent proof-of-principle studies demonstrate that combining experimental data with AlphaFold-like models, often with MSA manipulations, can improve static structure prediction and even show some proof of principles capturing conformational ensembles[129–135]. Such demonstrations highlight the potential of generative approaches; however, realizing this vision requires a more robust infrastructure to effectively integrate experimental datasets with ML methods, including substantial development to build robust infrastructure and better metrics that evaluate ML-derived ensembles against experimental ground truths rather than imperfect structural models, ensuring progress toward reliability and interpretability. We see the future being a tighter integration between machine learning and structural biology that can create a virtuous cycle: accelerating and improving experimental data processing and model building, while feeding back into more accurate ensemble predictors, recognizing that all structural data inherently encode ensembles and deserve to be modeled as such[103].

# Conclusion

The thermodynamic hypothesis from Christian Anfinsen states that “the native conformation is determined by the totality of interatomic interactions and hence by the amino acid sequence, in a given environment”[136]. Conformational ensembles are driven by the amino acid sequence and the environment, including any perturbations to the macromolecule. Here, we outline key conceptual and methodological foundations needed to more accurately predict these ensembles. Advancing our ability to predict conformational landscapes will expand our understanding of complex biological mechanisms and enable the development of novel therapeutic strategies. We can design small molecules that stabilize specific conformations by predicting low-population states or engineer antibodies whose binding depends on a precisely defined structural ensemble[137,138]. Likewise, successful enzyme design hinges on accurately predicting conformational ensembles, as catalysis inherently requires enzymes to traverse complex conformational landscapes[139,140].

It is also important to critically evaluate the current limitations of (static) structure prediction methods. Current limitations include achieving sub-angstrom accuracy or the "last Angstrom problem", and the accurate modeling of non-protein structures, including RNA. Theoretically, both issues may arise from the lack of ensemble representation. In proteins, many atoms undergo vibrations or subtle conformational fluctuations that are functionally relevant and may have evolved, while RNA is known to be inherently flexible[141]. However, it is also possible that ensemble prediction algorithms will perpetuate or make these problems worse. Further, existing methods still struggle to predict the effects of mutations, post-translational modifications, or other perturbations[142,143]. In principle, predicting ensembles could bridge this gap by linking perturbations to shifts in state populations, yet proof will require new data and benchmarks. Finally, many ensemble prediction methods are built off the architecture of static structure prediction models, which may not be optimal to capture conformational ensembles. Ensemble prediction will likely demand architectures that model distributions, not points.

Now is the time to build shared infrastructure for defining, collecting, modeling, encoding, and evaluating conformational ensembles. Such a framework would provide a powerful iterative loop of prediction, experimental validation, and refinement, bringing us closer to the ability to predict ensembles. This infrastructure is also necessary for extending conformational ensembles predictions to in-vivo context or obtaining timescale information. As high-throughput biophysical assays and in-vivo structure determination continue to provide new information on how ensembles change across conditions[144–146], we need infrastructure in place to incorporate this new data to enable active learning to improve the prediction of conformational ensembles in different environments and with different perturbations. Moreover, a unified ecosystem will pave the way for capturing kinetic properties, extending ensemble modeling from thermodynamic ensembles to temporally resolved protein dynamics. Overall, this shift promises to move structural biology beyond static snapshots toward a dynamic understanding of molecular behavior that captures the full complexity of biological systems.

# Acknowledgements

We thank Jared Sagendorf, Andrej Sali, Arthur Zalevsky, Frank Noe, and James Fraser for helpful feedback on this manuscript. S.A.W. is supported by the American Cancer Society. M.B. acknowledges funding from the European Research Council (ERC) under the European Union's Horizon 2020 research and innovation programme (Grant agreement No. 101086685 – bAIes).